\documentclass[10pt,letterpaper]{article}
\usepackage[top=0.85in,left=2in,footskip=0.75in]{geometry}

\usepackage{changepage}

\usepackage[utf8]{inputenc}

\usepackage{textcomp,marvosym}

\usepackage{fixltx2e}

\usepackage{amsmath,amssymb}

\usepackage{natbib}

\usepackage{nameref,hyperref}

\usepackage{microtype}
\DisableLigatures[f]{encoding = *, family = * }

\usepackage{rotating}

\usepackage{setspace} 
\raggedright
\usepackage[aboveskip=1pt,labelfont=bf,labelsep=period,justification=raggedright,singlelinecheck=off]{caption}

\makeatletter
\renewcommand{\@biblabel}[1]{\quad#1.}
\makeatother

\date{}

\usepackage{lastpage,fancyhdr,graphicx}
\usepackage{epstopdf}
\fancyheadoffset[L]{2.25in}
\fancyfootoffset[L]{2.25in}

\begin{document}
\vspace*{0.35in}

\begin{flushleft}
{\Large
\textbf\newline{A computational study on synaptic and extrasynaptic effects of
  astrocyte glutamate uptake on orientation tuning in V1}
}
\newline
\\
Konstantin Mergenthaler\textsuperscript{1},
Franziska Oschmann\textsuperscript{1},
Jeremy Petravicz\textsuperscript{2},
Dipanjan Roy\textsuperscript{1},
Mriganka Sur \textsuperscript{2},
Klaus Obermayer\textsuperscript{1,*}
\\
\bigskip
\bf{1}  Neural Information Processing Group, Fakult\"at IV and Bernstein Center for Computational Neuroscience, Technische
Universit\"{a}t Berlin, Berlin, Germany
\\
\bf{2} Picower Institute for Learning and Memory, Department of Brain and
Cognitive Sciences, Massachusetts Institute of Technology, Cambridge, MA
02139, USA
\\

\bigskip

* klaus.obermayer@mailbox.tu-berlin.de

\end{flushleft}
\section*{Abstract}
Astrocytes affect neural transmission by a tight control of the glutamate transporters which affect glutamate concentrations in direct vicinity to the synaptic cleft and in the extracellular space. The relevance of glutamate transporters for information representation has been supported by in-vivo studies in ferret and mouse primary visual cortex. A pharmacological block of glutamate transporters in ferrets broadened tuning curves and enhanced the response at preferred orientations. In knock-out mice with reduced expression glutamate transporters a sharpened tuning was observed. It is, however, unclear how focal and ambient changes in the glutamate concentration affect stimulus representation. Here, we developed a computational framework, which allows the investigation of synaptic and extrasynaptic effects of glutamate uptake on orientation tuning in recurrently connected network models with pinwheel-domain (ferret) or salt-and-pepper (mouse) organization. This model proposed that glutamate uptake shapes information representation when it affects the contribution of excitatory and inhibitory neurons to the network activity. Namely, strengthening the contribution of excitatory neurons generally broadens tuning and elevates the response. In contrast, strengthening the contribution of inhibitory neurons can have a sharpening effect on tuning. In addition, local representational topology also plays a role: In the pinwheel-domain model effects were strongest within domains - regions where neighboring neurons share preferred orientations. Around pinwheels but also within salt-and-pepper networks the effects were less strong. Our model proposes that the pharmacological intervention in ferret increases the contribution of excitatory cells, while the reduced expression in mouse increases the contribution of inhibitory cells to network activity.

\section*{Author Summary}
One of the key function of astrocytes is the clearance of neurotransmitters
released during synaptic activity. Its importance for stimulus representation in the cortex was hypothesized following experiments that showed changes
in selectivity when glutamate transport was blocked. Pharmacological and genetic interventions on glutamate transport considerably changed tuning width and strength of response in primary visual cortices of ferret and mouse.  Here,
we construct a modeling framework for visual cortices with pinwheel-domain and
salt-and-pepper-organizations, which allows the detailed investigation of
effects of altered glutamate uptake on orientation tuning. Our model proposes
that changes in the representation of stimuli gets less selective if changes
in glutamate uptake elicit stronger contribution of excitatory neurons to the
network activity and selectivity is sharpened for a higher contribution of
inhibitory neurons.

\section*{Introduction}

Over the last years the view on astrocytes changed from mere supporting tissue providing metabolic support to active partners in information transmission and processing
\cite{DePitta2012,Alvarellos-Gonzalez2012,Nadkarni2008,Reato2012,Perea2009}.
Strongest drive to this shift of the perspective was the development of calcium sensitive dyes \cite{Grynkiewicz1985} and the improvement of two-photon imaging \cite{Helmchen2005}. These technique allowed the simultaneously observation of calcium transients in both astrocytes and neurons {\em in-vivo} \cite{Schummers2008}. Several pathways have been identified how neuronal and synaptic activity drive astrocyte activity \cite{Perea2014a,Haydon2014,Benediktsson2012} or vice versa \cite{Perea2009,Araque2014,Chen2012,Nedergaard2012a}. Some of these pathways contain signaling cascades consisting of metabotropic receptors at the astrocyte membrane, internal second messenger
signaling and vesicular release from astrocytes
\cite{Panatier2011,Araque2014,DePitta2011}. Other pathways contain transporters and pumps in the astrocyte plasma membrane, which directly link neuron and astrocyte activity via control of ion- and
transmitter concentrations in a shared extracellular space \cite{Larsen2014,Rose2013}.
\newline
An {\em in-vivo} study in the ferret visual cortex (V1) revealed the relevance of astrocytes for stimulus representation in the cortex \cite{Schummers2008}. This study investigated the effects of a pharmacological block of the glutamate transport on the well-defined response to differently oriented gratings. While blocking the glutamate uptake in astrocytes leads to a stronger but less orientation selective response in neurons, the activity in astrocytes and the intrinsic optical signal were strongly attenuated. Another study revealed that a strongly reduced concentration of the primary astrocyte transporter (GLT-1) caused a sharpened orientation tuning \cite{Petravicz2014a}.
\newline
In a review \cite{Scimemi2009b} investigating how glutamate uptake might shape the synaptic glutamate concentration time course two key constraints were pointed out: geometry \cite{Rusakov1998} and transporter efficiency \cite{Diamond2001, Diamond2005,Thomas2011, Zheng2008}. First,
diffusion constraints, like a confined space \cite{Freche2011} and a clutter \cite{Min1998}, shape the glutamate concentration after the release. Particularly, the size and the geometry of synapses play a role in glutamate clearance \cite{Tarczy-Hornoch1998,Megias2001,Gulyas1999}. Moreover, it has been observed that glutamatergic synapses to
excitatory or to inhibitory cells differ in their geometry \cite{Koester2005a}. Therefore, glutamatergic synapses are considered as a determining factor for these two types of synapses \cite{Barbour1994}. In addition, Monte-Carlo modeling studies confirmed spatial constraints as a key determinant to glutamate clearance \cite{Freche2011,Rusakov2001,Barbour2001}. The second key constraint for the glutamate concentration time course are glutamate transporters, which shape the glutamate clearance from the synaptic cleft by buffering and complete uptake \cite{Danbolt2001, Scimemi2009}. While some transporter subtypes are also found on pre- and postsynaptic neurons \cite{Danbolt2001,Divito2014a}, the most abundant transporter (GLT-1) is highly concentrated on astrocyte processes ensheathing synapses \cite{Chaudhry1995,Benediktsson2012,Rusakov2014}. Dynamic changes in diffusion constraints occur primarily during maturation \cite{Thomas2011,Diamond2005}, but changes in neurotransmitter uptake can also be achieved by pharmacological blocking \cite{Schummers2008}, or genetic ablation \cite{Petravicz2014a}. Different effects of blocking glutamate transport on glutamate clearance have been found. One study proposes a shortening of glutamate clearance from the synaptic cleft when TBOA is applied, since less transporters are available to buffer glutamate within the cleft \cite{Scimemi2009}. Other studies propose a prolongation of the glutamate time course within the synapse during a block of the glutamate transport \cite{Murphy-royal2015, Barbour1994, Tong1994}. The modified glutamate concentration time course affects neurons via AMPA and NMDA receptors \cite{Tsukada2005, Bentzen2009}. During a blocked glutamate transport with TBOA a prolongation of AMPA-receptor mediated currents and a prevention of receptor desensitization were observed  \cite{Mennerick1999}. Moreover, high concentrations of TBOA lead to a self-sustained pathologic rapid firing or to cell-death \cite{Tsukada2005,Rothstein1996}. 
\newline
Based on the studies named above we hypothesize that glutamate
transporters shape physiological responses. The representation of stimulus specific features within the neo-cortex is largely considered to occur in networks which contain strong lateral connections with tightly calibrated excitatory and inhibitory contributions \cite{Stimberg2009,Shushruth2012,Marino2005}. This lead us to the question whether there are physiological
properties of the glutamate uptake which could elicit changes in the proportion of the excitatory and inhibitory contribution. 
Moreover, the proportion of excitatory and inhibitory contribution crucially depends on the difference in strength between excitatory and inhibitory neurons. Therefore, differences in extrasynaptic NMDA receptor-expression on excitatory and inhibitory neurons would determine the susceptibility of the network to ambient glutamate rise. To our knowledge studies which found differences in NMDA receptor properties on excitatory and inhibitory cells \cite{Martina2003,Martina2013} did not explicitly investigate extrasynaptic NMDA receptors. As such differences affect the proportion of excitatory and inhibitory contribution and no detailed experimental observations are available we incorporates differences in sensitivity into the model and explore its contribution to stimulus representation.
\newline
In the following, we first investigate changes in the glutamate decay
time within isolated synapses which comprise kinetic models for AMPA- and
NMDA-receptors. Similar to \cite{Allam2012,David2009} we investigate changes in the glutamate decay time depending on the fraction of open AMPA and NMDA receptors. We particularly focus on fractions of open receptors when stimulations follow Poisson processes with different rates. In a next step these detailed synapses are integrated in a 2D- network for ferret visual cortex. When
the glutamate transport is unchanged the network operates in a regime with strong lateral inhibitory and excitatory drive. For the integrated model we ask whether we can find combinations of glutamate decay times for synapses with excitatory and inhibitory connections which generate a similar loss in selectivity as in Schummers et al. \cite{Schummers2008}. As a second investigation we examine whether differences in the sensitivity to ambient glutamate between excitatory and inhibitory neurons shape orientation tuning in the network model. Motivated by the experiments which compare orientation tuning in GLT- wild type and knock-out mice we investigate the effects of different glutamate decay times and of different sensitivities to ambient glutamate. These experiments were also performed in a network with salt-and-pepper organization \cite{Runyan2013}.


\section*{Results}

\subsection*{Glutamate uptake and its effect on the transmission properties of single excitatory synapses}

\begin{figure}[!ht]
\begin{center}
\includegraphics[width=5in]{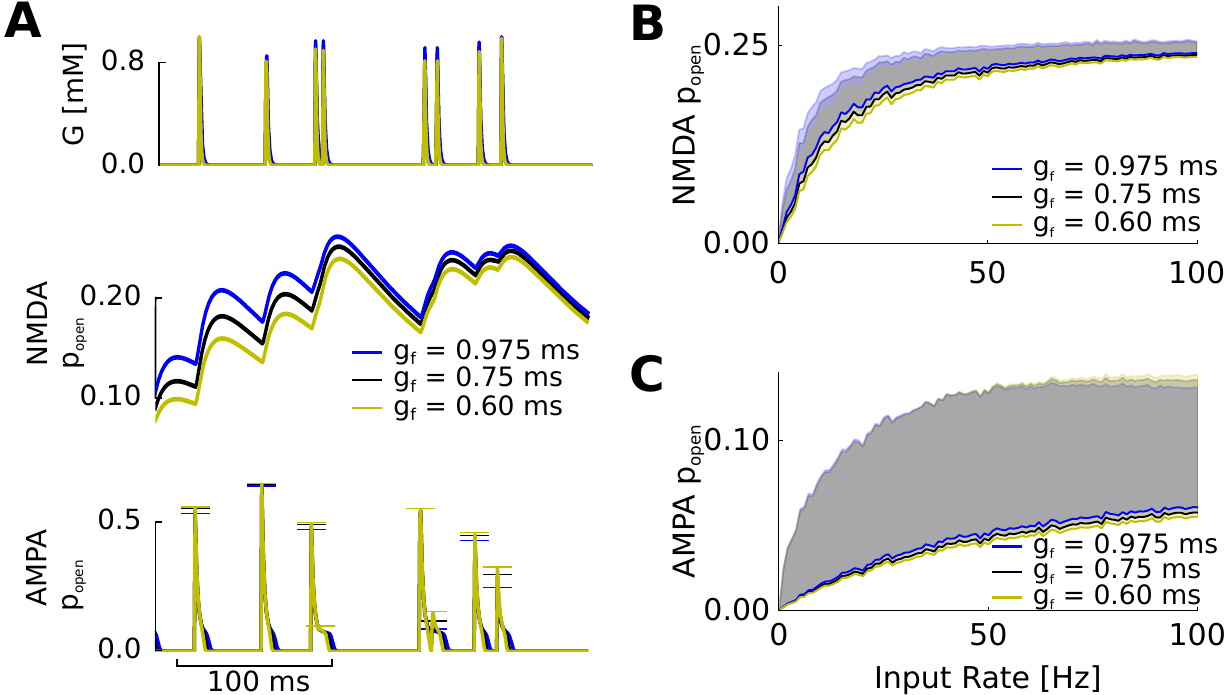}
\end{center}
\caption{
{\bf Simulation of a single synapse. A} (upper) Time course of the glutamate concentration $G$ for different decay times g\textsubscript{f} (fast: $0.6$~ms (yellow); base line: $0.75$~ms (black); slow: $0.975$~ms (blue)). The glutamate pulses are generated by a Poisson-rate of $40$~Hz, which drive the NMDA \& AMPA
receptors. (middle) Time course of the fraction of open NMDA receptors for different glutamate decay time constants following the glutamate pulses shown in the upper figure. (lower) Time course of the fraction of open AMPA receptors for different glutamate decay time constants following the glutamate pulses shown in the upper figure. {\bf B} Fraction of open NMDA receptors after stimulation of $2$~s with
different Poisson-rates. Bold lines show the average proportion of open receptors and the
shaded area its standard deviation. Largest differences in mean and strongest
variation are found around $15$~Hz. For high rates differences vanish. {\bf C}
Fraction of open AMPA receptors after stimulation of $2$~s with
different Poisson-rates. Bold lines show the average proportion of open receptors and the
shaded area its standard deviation. Average proportion of AMPA-receptors increase with
rate and decay-constants. Standard deviation is largest and less rate dependent
for short decay times. \label{fig:synapses}
}
\end{figure}

We studied the influence of astrocyte-mediated glutamate uptake on the transmission
properties of glutaminergic synapses in a simplified setting.
Here, the dynamics of synaptic AMPA receptors and NMDA receptors were described using kinetic models with 3 and 5 states (see Methods: Neurotransmitter concentration
\& receptor dynamics). The glutamate concentration in the synaptic cleft was quantified by
bi-exponential pulses following every presynaptic spike. Different
glutamate decay time constants accounted for changes in the efficacy of astrocytic glutamate
uptake, where short (long) decay times corresponded to fast (slow)
glutamate uptake.
\newline
Fig.~\ref{fig:synapses} shows the fraction of open NMDA and AMPA receptors in response to Poisson-distributed spike trains for three different decay
time-constants of the glutamate pulses. The fraction of open NMDA receptors was mostly affected by different glutamate decay times when both the fraction of open NMDA receptors and the number of glutamate pulses were low (see Fig.~\ref{fig:synapses}A). As a consequence, different glutamate decay times had the biggest impact on the fraction of open NMDA receptors when the frequency of the presynaptic spike rates ranged between 10 and 15 Hz (see Fig.~\ref{fig:synapses}B). 
\newline
The fraction of open AMPA receptors was only marginally influenced by different glutamate decay times for large intervals between presynaptic spikes. However, an increase of the glutamate decay time prolonged the time to complete receptor closure (see Fig.~\ref{fig:synapses}A). Moreover, the effect of different glutamate decay time constants on the fraction of open AMPA receptors increased with the presynaptic spike rate (see Fig.~\ref{fig:synapses}B).

\subsection*{Effects in a V1 with pinwheel-domain organization}

By asking whether affecting glutamate transport might have effects on
representation of information in a recurrently connected networks, particular importance can be
attributed to mechanisms weighting the contribution of excitatory and
inhibitory populations. As the glutamate decay does not only depend on glutamate transporters but also on synapse geometries we independently varied
the glutamate decay time for lateral synapses to either excitatory
(EE-synapses) or inhibitory (IE-synapses) neurons and investigated changes in
tuning. In our single layer model lateral synapses were synapses formed between neurons within the layer in contrast to afferent synapses, which originate from lower layers. 

\begin{figure}[!ht]
\begin{center}
\includegraphics[width=5in]{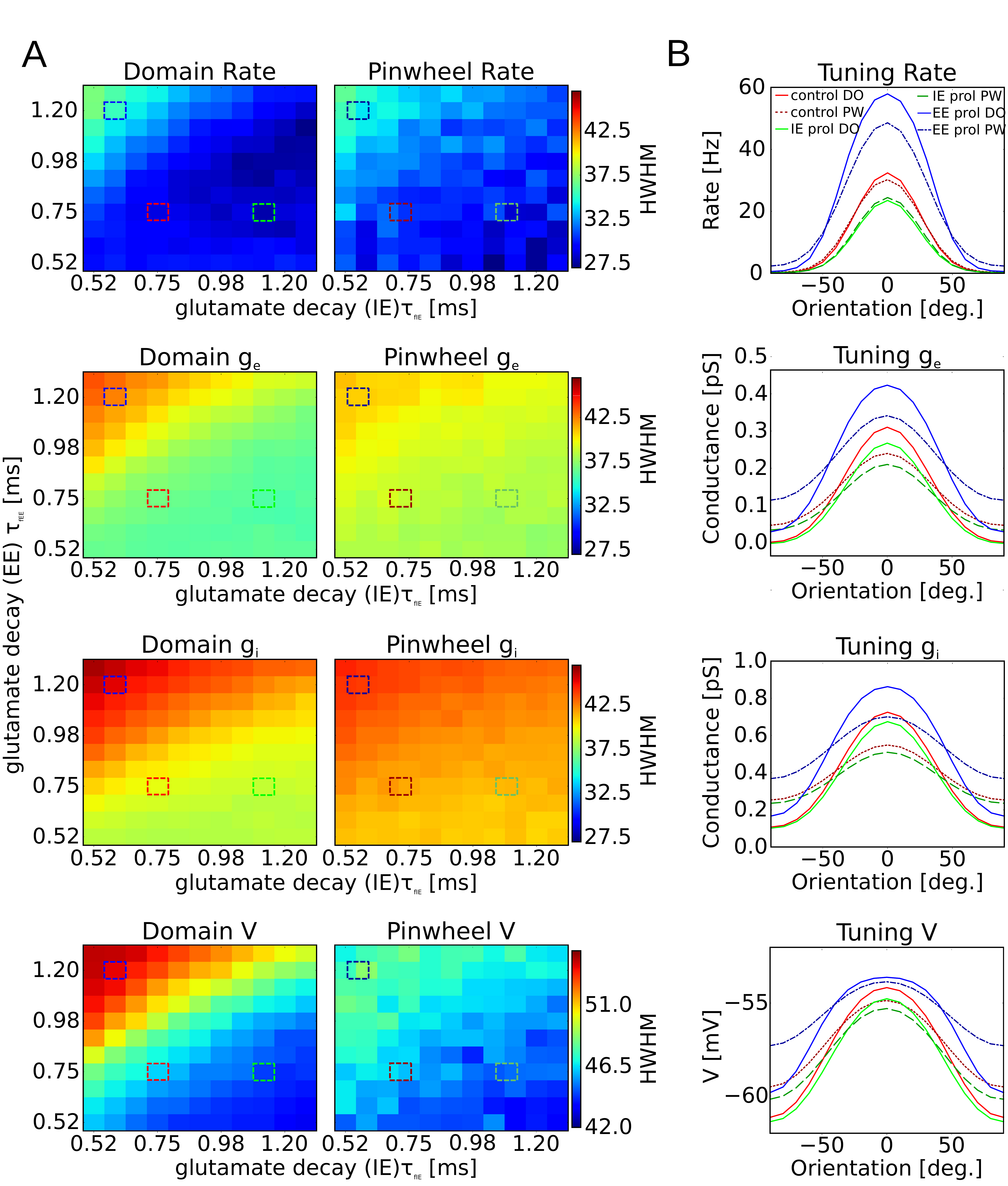}
\end{center}
\caption{
{\bf Synaptic effect in pinwheel-domain network model. A:} Glutamate decay
within the IE- (horizontal axis) and EE-synapses (vertical axis) are
independently varied. The reference condition point is $0.75$/$0.75$~ms (red
box). Values below $0.75$~ms are shortened and values
above $0.75$~ms are prolonged glutamate clearance
values. Half-width-at-half-max (HWHM) values (color coded) of the tuning
curves are separately derived for neurons within orientation domains (left) and neurons close to pinwheels
(right) for the firing rate, the received excitatory conductance, the received
inhibitory conductance, and the membrane potential in excitatory neurons. All
four investigated properties show a loss in selectivity and increased values
if prolongation of glutamate decay preferentially occurs in connections to
excitatory neurons. The effect is even stronger with a simultaneous reduction
in decay time for connections to inhibitory neurons (exemplary: blue box). If prolongation would
mostly occur in connections to inhibitory neurons responses are slightly
sharpened (exemplary: green box). {\bf B} Tuning curves for the exemplary
points from {\bf A}. }
\label{fig:ferret_glu_prol}
\end{figure}

\paragraph*{Synaptic mechanism}
Starting from our reference point with the
same decay constant ($0.75$~ms, red box in Fig.~\ref{fig:ferret_glu_prol}A)
for EE-synapses and IE-synapses we observed  that a prolongation
of the glutamate decay time within EE-synapses broadens the firing rate tuning (Half-width-at-half-max: HWHM increases). The reference point ($\tau_{fEE} = \tau_{fIE} = 0.75$~ms) was chosen in accordance with values derived in \cite{Diamond2005}. The broadening of tuning curves was even
stronger with a simultaneous reduction of the decay constant in IE-synapses (exemplary point: blue box in
Fig.~\ref{fig:ferret_glu_prol}A). Slight sharpening was observed
when prolongation occurs mostly within IE-synapses (reference point: green box in
Fig.~\ref{fig:ferret_glu_prol}A). This picture held within domain
centers as well as close to pinwheels. However, close to pinwheels we observed markedly smaller effects for different decay constants. 
\newline
In addition to
changes in firing rate tuning a very similar picture was found for excitatory
and inhibitory conductances as well as the sub-threshold membrane
potential. Interestingly, the membrane potential showed a prominent
sharpening when the glutamate decay time within IE-synapses was prolonged
(Fig.~\ref{fig:ferret_glu_prol}A lower-left panel). The
broadening of the response for a prolonged decay in EE-synapses and a shortened decay in
IE-synapses went hand in hand with an increase in firing rates (Fig.~\ref{fig:ferret_glu_prol}B). Therefore, a
detailed prolongation of the glutamate decay time within EE-synapse and a simultaneous
reduction of the glutamate decay time within IE-synapses provided a plausible condition for
the experimentally observed change in tuning response during pharmacological block of the glutamate
transport in ferret V1 \cite{Schummers2008}.

\begin{figure}[!ht]
\begin{center}
\includegraphics[width=5in]{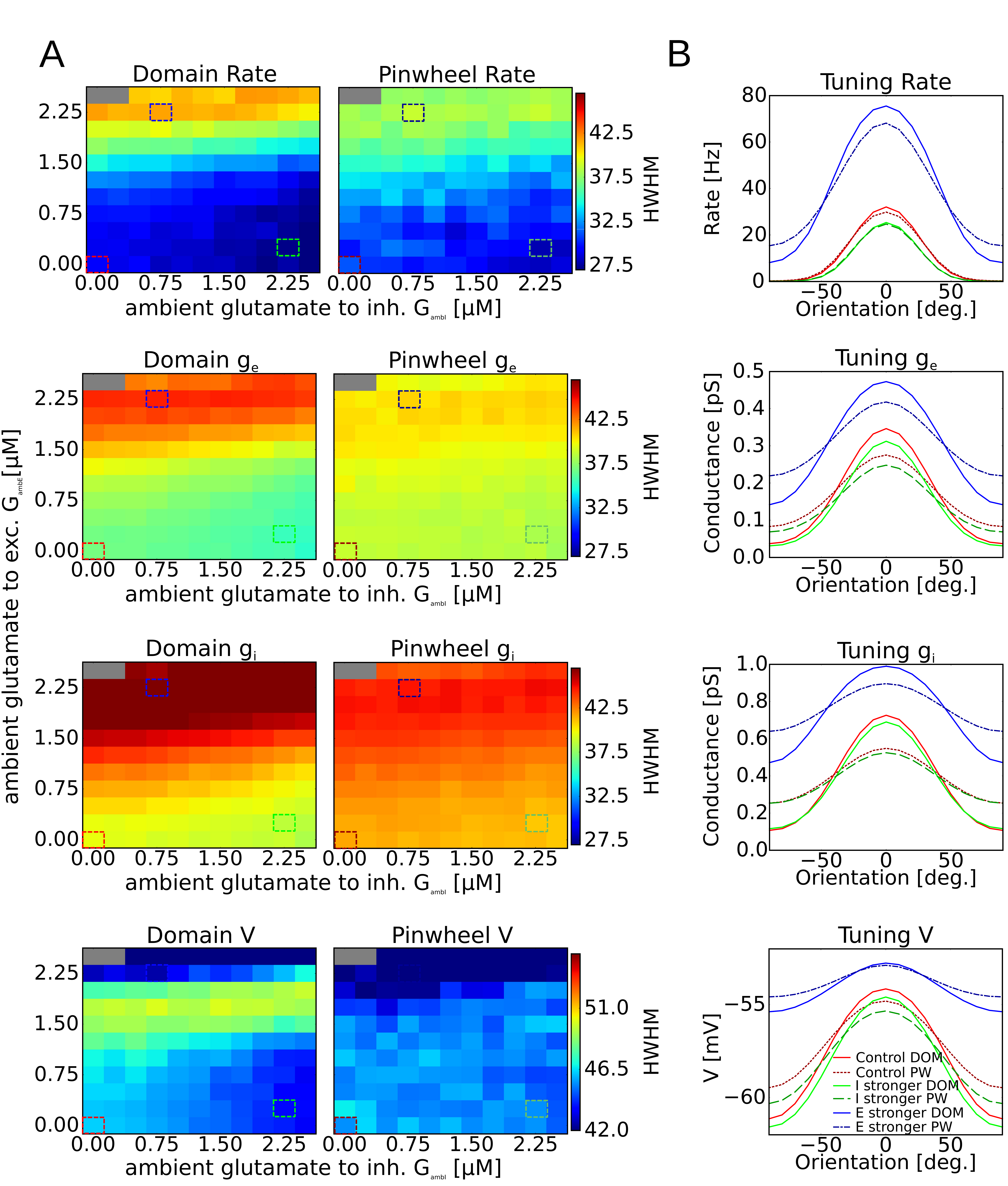}
\end{center}
\caption{
{\bf Extrasynaptic effect in pinwheel-domain network model. A} Different elevated levels
of ambient glutamate sensed by excitatory (vertical axis) and inhibitory
(horizontal axis) neurons represent different efficiancies of NMDAR on
excitatory and inhibitory neurons. Half-width-at-half-max (HWHM) values (color
coded; gray = self sustained network activity) are derived for neurons within domains (left) and close to pinwheels (right) for firing rate, excitatory and inhibitory conductance and
membrane potential. An increase in NMDAR-currents to excitatory neurons
(exemplary: blue box) reduces orientation tuning selectivities and generally
increases responses. An increase in NMDAR-currents on inhibitory neuron give
rise to slightly sharpened but weaker responses (exemplary: green box). {\bf B}
Exemplary tuning curves from {\bf A}.}
\label{fig:ferret_amb_glu}
\end{figure}

In addition to changes in lateral connections,
a simultaneous change in afferent excitatory connections (EA-synapses: to
excitatory neurons, and IA-synapses: to inhibitory neurons) might occur. The
exploration of the simultaneous prolongation in EE- and EA-synapses as well
as in IE- and IA-synapses revealed, that prolongation and shortening became more
effective and enhance the strengthening effect of one population above the other. However, no qualitative change were observed (data not shown).
\paragraph*{Extrasynaptic mechanism}
Another mechanism that weights the contribution of the excitatory and the inhibitory population differently and could originate from changes in the glutamate transport is a difference in sensitivity to the ambient glutamate level of excitatory and inhibitory neurons via extrasynaptic NMDARs. As a proxy for
different NMDAR-densities we independently varied the ambient glutamate concentration affecting NMDARs on excitatory and inhibitory cells. With an increase of the ambient glutamate concentration effective on the excitatory
neurons the orientation tuning broadened (higher HWHM values, cf. Fig.~\ref{fig:ferret_amb_glu}A). In addition, responses at preferred and non preferred orientations increased
(Fig.~\ref{fig:ferret_glu_prol}B). Again this effect was much more pronounced within domains and much weaker around pinwheels. For an even stronger effect on excitatory neurons the network entered a state of pathological self sustained activity. The effect of NMDAR-currents on inhibitory neurons was
again small and only small changes in the tuning width were observed.
    
\subsection*{Effects in a V1 with salt-and-pepper organization}

The smaller effects around pinwheels called for the investigation of effect of
glutamate decay times in a network with a salt-and-pepper organization. We investigated the effect of differential changes of the glutamate decay time in a
model, which was calibrated to reproduce the observed firing rate tuning in mouse V1 \cite{Runyan2013}. 

\begin{figure}[!ht]
\begin{center}
\includegraphics[width=5in]{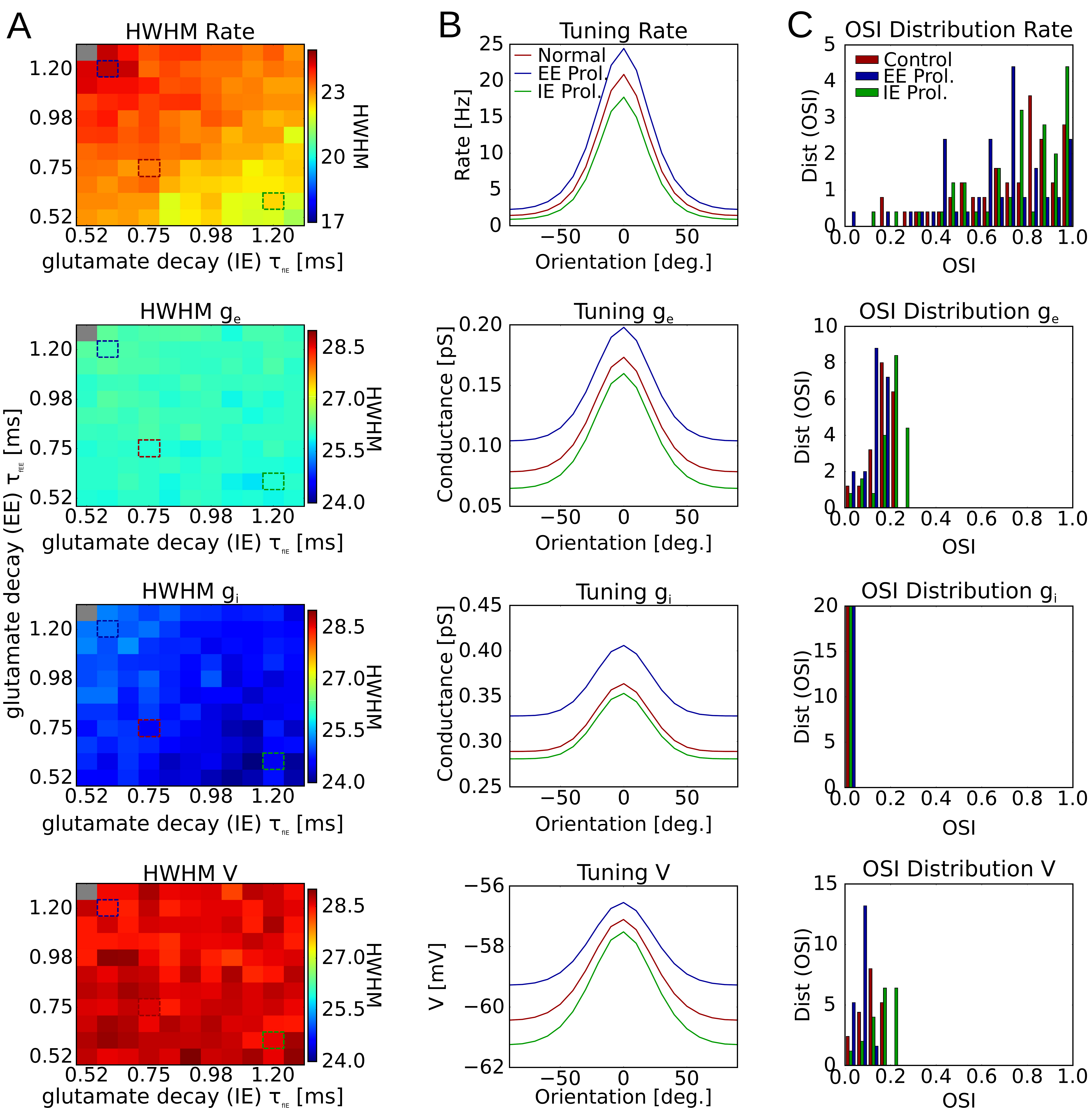}
\end{center}
\caption{
{\bf Synaptic effect in salt-and-pepper network model. A:} Glutamate decay
time is separately varied in lateral synapses to excitatory neurons (vertical axis)
and inhibitory neurons (horizontal axis) and HWHM of tuning curves for firing
rate, excitatory and inhibitory conductance, and membrane potential are shown
color-coded. Only HWHM for rate shows a prominent effect of changes in glutamate
decay time. Boxes are for exemplary points (reference: red; prolongation in
connections to excitatory neurons (EE): blue; prolongation to inhibitory
neurons (IE): green) {\bf B} Tuning curves for exemplary points show upward (prolonged EE)
and downward (prolonged IE) shifts with little change in tuning width. {\bf C}
Orientation-Selectivity (OSI)-distributions for the exemplary points show
higher OSI-values when IE-decay is prolonged (green), and lower OSI-values
when EE-decay is prolonged (blue) for rate, excitatory conductance and
membrane potential.}
\label{fig:mouse_glu_prol}
\end{figure}

\paragraph*{Synaptic mechanism}
It turned out that changes in the glutamate decay constants
only weakly changed the firing rate orientation tuning and had only negligible
effects on the other quantities when considering half-width-at-half-max
values (HWHM) Fig.~\ref{fig:mouse_glu_prol}A. While the shape of the tuning curves hardly changed and the tuning
curves were mostly shifted upward for prolonged glutamate decay in EE-synapses and shifted
downward for prolonged glutamate decay in IE-synapses
Fig.~\ref{fig:mouse_glu_prol}B, responses at preferred and non-preferred orientations changed. The
orientation-selectivity-index (OSI), however, merged shift- and shape-changes and OSI-distributions were either shifted to lower values when glutamate decay
was prolonged in EE-synapses or were shifted to higher values when glutamate decay was prolonged in IE-synapses. For strongly prolonged glutamate decay times in EE-synapses with simultaneous shortening in IE-synapses the network reached self-sustained firing.

\begin{figure}[!ht]
\begin{center}
\includegraphics[width=5in]{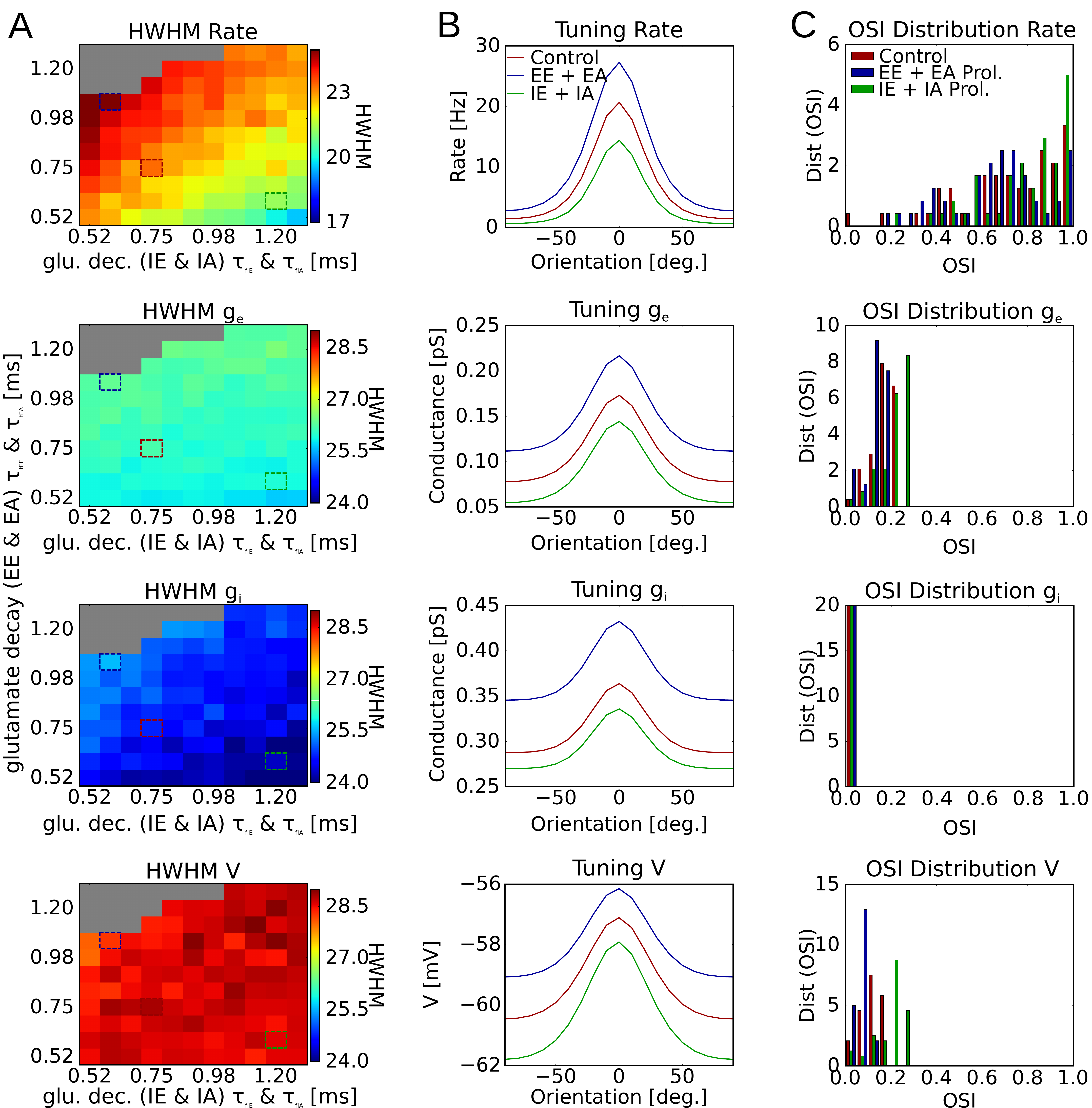}
\end{center}
\caption{
{\bf Synaptic effect in salt-and-pepper network model -- all synapses. A:} In
contrast to the exploration in Fig.~\ref{fig:mouse_glu_prol} the decay time in
afferent synapses is varied alongside the lateral ones. Again the HWHM for
rate shows broadening for prolonging EE-synapses and EA-synapses and some
sharpening for prolonged IE- and IA-synapses. In addition small difference
could also be found in the sub threshold properties. Boxes are for exemplary points (reference: red; prolongation in
connections to excitatory neurons (EE + EA): blue; prolongation to inhibitory
neurons (IE + IA): green) {\bf B} The tuning curves for exemplary points show
that changes in HWHM are minor in comparison to the strong shifts (upward for
EE + EA-synapses and downward for IE + IA-synapses) {\bf C} The
OSI-distribution combining baseline-shifts and width-changes show clearer
separation of selected points. Generally, lower OSI-values are observed when
glutamate decay in EE + EA-synapses is prolonged (blue) and higher OSI-values if
prolongation occurs mostly in IE + IA-synapses (green).}
\label{fig:mouse_all_glu_prol}
\end{figure}

Again during a simultaneous prolongation in EE- and EA-synapses as well as in IE- and
IA-synapses the effect on HWHMs were more pronounced
(Fig.~\ref{fig:mouse_all_glu_prol}A). Now, the selectivity loss for a prolonged decay in EE- and EA-synapses and a shortened decay in IE- and IA-synapses, and the selectivity increase for a prolonged decay in IE- and
IA-synapses and a shortened decay in EE- and EA-synapses, were also visible in the HWHMs of the sub threshold
properties. Nevertheless, the biggest change occurred as upward or downward shifts of the tuning
curves independent of the preferred orientation
(Fig.~\ref{fig:mouse_all_glu_prol}B). Both changes were reflected in changes in
OSI-values and prolongation to excitatory neurons shifted OSI-distributions to
lower values, and prolongation to inhibitory neurons shifted OSI-distributions
to higher values (Fig.~\ref{fig:mouse_all_glu_prol}C). 

\begin{figure}[!ht]
\begin{center}
\includegraphics[width=5in]{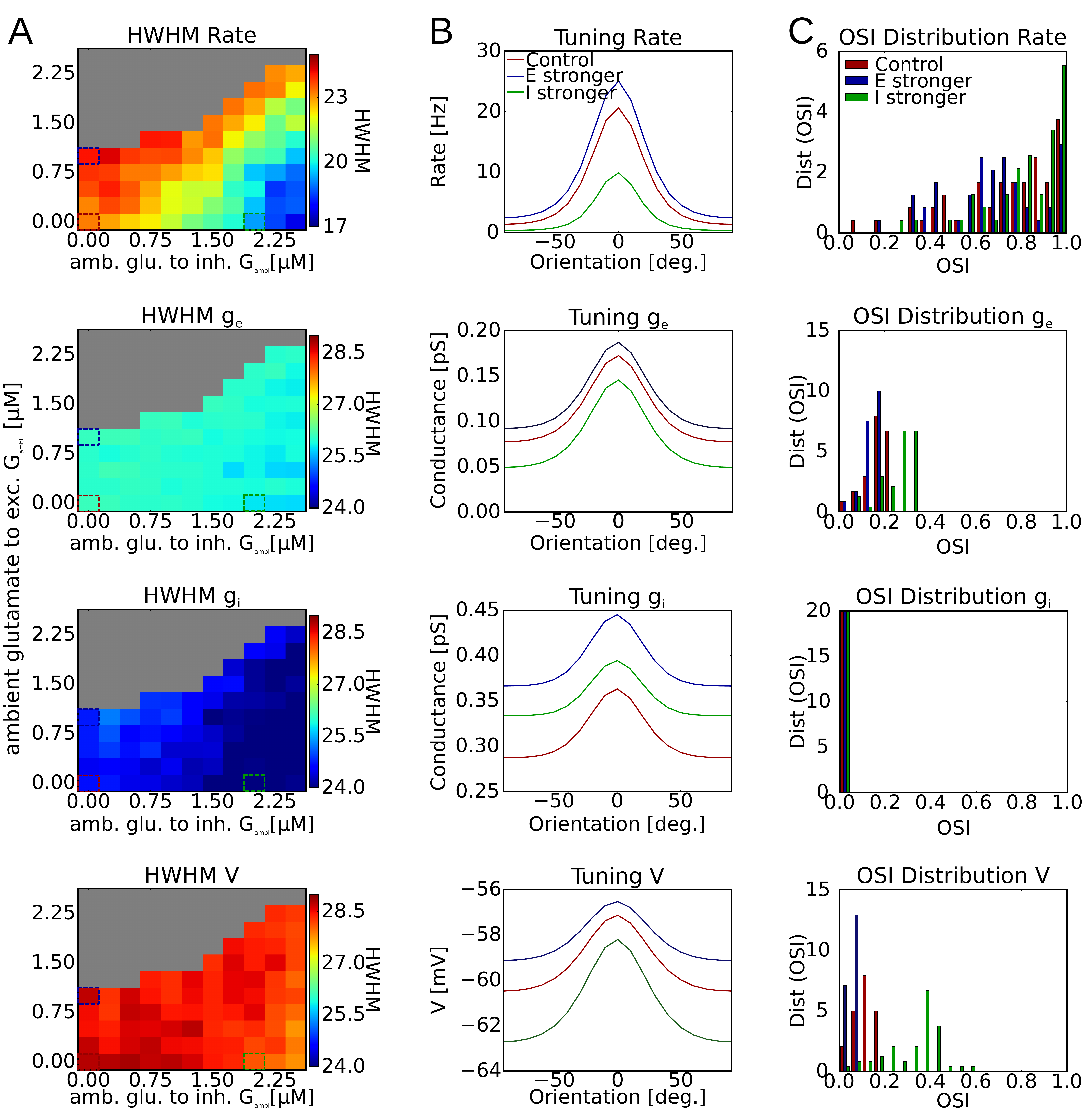}
\end{center}
\caption{
{\bf Extrasynaptic-effect in salt-and-pepper network model. A:} Again
different elevated levels of ambient glutamate to excitatory (vertical axis)
and inhibitory (horizontal axis) concentrations of ambient glutamate represent
different NMDAR efficiancies. Tuning (HWHM) gets less selective if extrasynaptic NMDAR-currents
have a stronger effect on excitatory neurons and more selective if NMDAR-currents
are higher on interneurons in firing rates and again much weaker in the other
variables. {\bf B} The tuning curves for sub-threshold properties show strong
stimulus-orientation independent changes. {\bf C} OSI-distributions show
higher values when the inhibitory population is primary target of ambient
glutamate and lower values when the excitatory population is primary target.}
\label{fig:mouse_amb_glu}
\end{figure}

\paragraph*{Extrasynaptic mechanism}
We used different ambient glutamate
concentrations as a proxy for different sensitivities of inhibitory and
excitatory neurons to elevated ambient glutamate. 
In a salt-and-pepper network we observed that stronger
sensitivity of excitatory neurons broadened the tuning
(Fig.~\ref{fig:mouse_amb_glu}A blue box). For inhibitory
cells more sensitive to ambient glutamate the HWHMs of the tuning curves of the firing rates and the conductances were markedly
reduced (Fig.~\ref{fig:mouse_amb_glu}A green
box). The membrane potential showed almost unchanged HWHM-values. For the synaptic mechanism the strongest effect were orientation
independent shifts (Fig.~\ref{fig:mouse_amb_glu}B), which elevated the baseline
activity for ambient glutamate mostly affected excitatory cells, and pulled down
the baseline values for ambient glutamate mostly affected inhibitory
cells. The prominent baseline shifts and the changes in tuning width combined
to clear shifts in the OSI-distributions (Fig.~\ref{fig:mouse_amb_glu}C). For more sensitive
inhibitory neurons OSI-distributions were shifted to higher
values and for more sensitive excitatory neurons OSI-distributions were shifted to smaller values. 

\section*{Discussion}
Schummers et al. \cite{Schummers2008} observed a loss in selectivity to oriented
gratings when glutamate transport is blocked pharmacologically. Our presented model
reproduces the loss in selectivity, but only if changes in the glutamate transport
enhance the contribution of the excitatory population. Such a strengthening was achieved
via two different pathways for glutamate uptake. One was the prolongation of glutamate 
decay within synapses to excitatory neurons. The second one was a postulated higher 
sensitivity of excitatory neurons to ambient glutamate. The importance of
shifts in excitatory vs inhibitory contribution mediated by changes in
glutamate transport can even be seen in a model with a salt-and-pepper
organization of preferred orientations. Such a model 
shows sharpened tuning (higher OSI-values) as for GLT-1$^{+/-}$-mice in \cite{Petravicz2014a}, 
but only if the contribution of inhibitory neurons is strengthened. Interestingly, 
the pinwheel-domain network showed an interaction between mapOSI as well as synaptic and extrasynaptic glutamate uptake effects. Changes in tuning were always stronger in domains
and much less pronounced close to pinwheels. Following, the analogy of neurons within a salt-and-pepper
network as neurons at pinwheels in a pinwheel-domain network it is not unexpected that effects on
tuning width (HWHM) are small in such a network. Particularly the observation
of lower effects of changes in lateral connections onto HWHM values but
orientation unspecific shifts of tuning curves is in line with a suggested
stronger contribution of weakly tuned inhibitory neurons as in
\cite{Bopp2014}. Interestingly, the
salt-and-pepper network with fewer lateral connections is more susceptible to pathologic self-sustained firing.   
For the pinwheel-domain as well as for the salt-and-pepper map we achieved to directly link 
effects of glutamate uptake to changes in information representation. To link these we, however, 
were forced to construct rather complex models with a lot of fixed parameters and a lot of detail. 
\newline
We took deliberate care in selecting fixed parameters to be in a physiological range, e.g., 
parameters describing the Hodgkin-Huxley dynamics of single neurons stem from 
models largely used for neurons in visual cortex \cite{Stimberg2009,Marino2005,Destexhe2001}. The necessity of 2D-network structures with 
local lateral connections follows arguments in 
\cite{Stimberg2009,Marino2005} and \cite{Roy2013arxiv}. It allowed us to calibrate the model in 
reference condition to match experimentally observed orientation tuning in pinwheel-domain networks
 \cite{Marino2005} and salt-and-pepper networks \cite{Runyan2013}
and to investigate interactions of local heterogeneity in representing stimulus features with effects of 
glutamate uptake. With the low number of connections and neurons and the very high
peak synaptic conductances we assume that each neuron and connection is representative for a 
subpopulation sharing exactly the same features as the representative single neuron. 
\newline
In contrast to our separate investigations of synaptic and extrasynaptic effect, 
we expect that glutamate uptake experiments show a combined effect of both mechanisms. 
To confine their exact contributions within the cortex, or answer whether changes in synaptic 
clearance or raising ambient glutamate can be ruled out -- due to no effective shift towards excitation or 
inhibition, required new careful experiments. For the synaptic effect, first synapses onto excitatory
and inhibitory cells need to be separately investigated and separate assessments of synapses geometry, size,
and transporter densities is required. Second, in single synapse studies -- either experimental or detailed
modeling studies similar to \cite{Scimemi2009b,Rusakov1998,Freche2011} -- effects on glutamate clearance and the susceptibility 
to altered glutamate uptake
for the two types of synapses need to be investigated. For the extrasynaptic mechanism  a separate 
estimation of only extrasynaptic NMDA receptor densities on excitatory and inhibitory cells 
would allow to estimate the effective impact of ambient glutamate. 
\newline
For the single synapse models we observed interactions between glutamate clearance decay time and firing
rate in the contribution to average and fluctuations in open fractions of receptors. This leads to a 
range of medium frequencies (10-15~Hz) where NMDA-receptors show the highest sensitivity to changes
in glutamate clearance. Similarly, AMPA-receptor fluctuations are most sensitive in a similar range.
Considering that
complex synapses will be present in networks which transit between fluctuation and mean driven phases \cite{Litwin-Kumar2012,Renart2007}, we 
propose that changes in glutamate clearance interact with the cortical dynamical state.
\newline
Finally, in the context of astrocytes as active partners, a short coming of our model is that both 
pathways of glutamate uptake were investigated without intrinsic dynamics. Further investigations on
the effects of glutamate uptake in networks would largely benefit from coupled dynamic models of
neurons and astrocytes. In such models the dynamic intrinsic state of an astrocytes, e.g. Ca$^{2+}$-concentration
would interact with glutamate uptake and finally the neighboring neurons.

\section*{Methods}
\subsection{Neuron model and postsynaptic currents}
\paragraph*{Concentration of neurotransmitter in the synaptic cleft and channel kinetics.}
$G_Y$ describes the time course of the neurotransmitter concentration in the synaptic cleft for the the excitatory neurotransmitter glutamate ($G_E$) and the inhibitory neurotransmitter GABA ($G_I$). The time course of the neurotransmitter concentration in response to a presynaptic action potential follows a bi-exponential function:
\[
G_{Y}(t) = \frac{1}{\tau_{fY} - \tau_{rY}} \sum_{t_k<t} \left( \exp \left(- \frac{t_k
  - t}{\tau_{fY}}\right) - \exp \left( - \frac{t_k  -t}{\tau_{rY}} \right) \right).
\]
Here, the rise and decay constants $\tau_{rY}$ and $\tau_{fY}$ (r: rise, f: decay) vary for different pairings of the post- (left letter) and presynaptic (right letter) cell type ($Y \in \{EE,EI,IE,II\}$, E: excitatory, I: inhibitory). $t_k$ denotes the arrival time of the action potential. Parameter values are summarized in Table~\ref{tab:ligand}. We chose $\tau_{rEE} = \tau_{rIE} = \tau_{rE}$, $\tau_{rEI} = \tau_{rII} = \tau_{rI}$, and $\tau_{fEI} = \tau_{fII} = \tau_{fI}$. The rise constant $\tau_{rE}$ remained fixed, because of its small value. Variations in the decay constants $\tau_{fEE}$ and $\tau_{fIE}$ accounted for changes in the astrocytic glutamate uptake. $G_{Y}$ was normalized, such that the peak concentrations of glutamate and GABA were set to $1$~mM \cite{Clements1992b,Vizi2010}.
\begin{table}[!ht]
\caption{
\bf{Ligand gated receptor dynamics}}
\begin{tabular}{llll}
Parameter & Value & Description & Source \\ 
\hline
\multicolumn{4}{l}{{\em Synaptic -- Neurotransmitter}}\\
$\tau_{rE}$ & $0.16$~ms & Glutamate concentration rise time &
\cite{Diamond2005} \\
$\tau_{fIE}$ & $0.545$--$1.275$~ms & Exc. to inh. concentration decay time & \cite{Diamond2005}* \\
$\tau_{fEE}$ & $0.545$--$1.275$~ms & Exc. to exc. concentration decay time & \cite{Diamond2005}*\\
$\tau_{rI}$ & $0.29$~ms & GABA concentration rise time & $\dagger$ \\
$\tau_{fI}$ & $0.291$~ms & GABA concentration decay time & $\dagger$ \\
\hline
\multicolumn{4}{l}{{\em Synaptic -- AMPA Channel dynamics}}\\
$R_{ar}$ & $0.065$~s$^{-1}$ & AMPAR resensitization rate & \cite{Saftenku2005} \\
$R_{ad}$ & $5.11$~s$^{-1}$ & AMPAR desensitization rate & \cite{Saftenku2005}\\
$R_{ao}$ & $25.39$~s$^{-1}$ & AMPAR opening rate & \cite{Saftenku2005} \\
$R_{ac}$ & $4.$~s$^{-1}$ & AMPAR closing rate & \cite{Saftenku2005} \\
$K_B$ & $0.44$~mM & AMPAR binding rate & \cite{Saftenku2005} \\
\hline
\multicolumn{4}{l}{{\em Synaptic -- NMDA Channel dynamics}}\\
$R_{nb}$ & $1 \times 10^6 $~M$^{-1} $s$^{-1}$ & NMDAR binding rate & \\
$R_{nu}$ & $12.9$~s$^{-1}$& NMDAR unbinding rate & \cite{Destexhe1998}\\
$R_{nd}$ & $8.4$~s$^{-1}$ & NMDAR desensitization rate & \cite{Destexhe1998}\\
$R_{nr}$ & $6.8$~s$^{-1}$ & NMDAR resensitization rate & \cite{Destexhe1998}\\
$R_{no}$ & $46.5$~s$^{-1}$ & NMDAR opening rate & \cite{Destexhe1998}\\
$R_{nc}$ & $73.8$~s$^{-1}$ & NMDAR closing rate & \cite{Destexhe1998}\\
\hline
\multicolumn{4}{l}{{\em Synaptic -- GABA$_A$ Channel dynamics}}\\
$R_{gb1}$ & $20 \times 10^6 $~M$^{-1} $s$^{-1}$ & GABA$_A$R binding rate 1& \cite{Destexhe1998}\\
$R_{gb2}$ & $10 \times 10^6 $~M$^{-1} $s$^{-1}$ & GABA$_A$R binding rate 2 & \cite{Destexhe1998}\\
$R_{gu1}$ & $4.6 \times 10^3$~s$^{-1}$& GABA$_A$R unbinding rate 1 & \cite{Destexhe1998}\\
$R_{gu2}$ & $9.2 \times 10^3$~s$^{-1}$& GABA$_A$R unbinding rate 2 & \cite{Destexhe1998} \\
$R_{go1}$ & $3.3 \times 10^3$~s$^{-1}$ & GABA$_A$R opening rate 1& \cite{Destexhe1998}\\
$R_{go2}$ & $10.6 \times 10^3$~s$^{-1}$ & GABA$_A$R opening rate 2& \cite{Destexhe1998}\\
$R_{gc1}$ & $9.8 \times 10^3$~s$^{-1}$ & GABA$_A$R closing rate 1& \cite{Destexhe1998}\\
$R_{gc2}$ & $410 $~s$^{-1}$ & GABA$_A$R closing rate 2 & \cite{Destexhe1998} \\
\end{tabular}
\begin{flushleft}*A range of values around 0.75ms (derived in \cite{Diamond2005a}) was explored. 
$\dagger$ Rise and decay constants were chosen such that the mean squared distance between the bi-exponential function and the concentration of GABA as a function of time calculated as in \cite{Destexhe1998} was minimal (particle swarm optimization).
\cite{Destexhe1998} 
\end{flushleft}
\label{tab:ligand}
\end{table}
%
Fig.~\ref{fig:kinetic_schemes} shows the kinetic schemes used for the AMPA-, NMDA-, and GABA$_A$-channels. The AMPA-channel is described by one closed, one desensitized and one open state \cite{Saftenku2005}. The NMDA-channel passes through three closed, one desensitized and one open state \cite{Lester1992}. The GABA-A channel has three closed and two open states \cite{Destexhe1998}.

\paragraph*{Neuron model} Neurons are described by conductance-based point neuron models,
where changes of the membrane voltage $V_X$ for excitatory ($X=E$)
and inhibitory ($X=I$) neurons are driven by a sum of transmembrane currents:
\begin{equation}
C_m \frac{dV_X}{dt} = - I_{L,X} - I_{int,X} - I_{syn,X} - I_{amb,X} - I_{bg,X}.
\end{equation}
$C_m$ denotes the membrane capacitance and $t$ the time. We consider:
(i) a leak current $I_{L,X} = -g_{L,X} (V_X - E_L)$ with leak conductance 
$g_{L,X}$ and reversal potential $E_L$, (ii) a sum $I_{int,X}$ of three
Hodgkin-Huxley type voltage-gated intrinsic currents (see below), (iii)
the total synaptic ligand-gated current $I_{syn,X}$, (iv) a ligand-gated current
$I_{amb,X}$ driven by extrasynaptic glutamate, and (v) a background
current $I_{bg,X}$ for inducing a realistic level of spontaneous activity. Parameters are
summarized in
Table~\ref{tab:neuron}.
\begin{table}[!ht]
\caption{
\bf{Membrane capacitance and parameters for the leak, intrinsic, and background currents of the neuron model.}}

\begin{tabular}{llll}
Parameter & Value & Description & Source \\ 
\hline
\multicolumn{4}{l}{{\em Membrane capacitance and leak current}}\\
$C_m$ & $0.35$~nF & Membrane capacitance & \cite{Schummers2007} \\
$g_{L,E}$ & $15.7$~nS & Leak conductance, excitatory neurons & \cite{Schummers2007} \\
$g_{L,I}$ & $31.4$~nS & Leak conductance, inhibitory neurons & \cite{Schummers2007} \\
$E_{L}$ & $-80$~mV & Reversal potential & \cite{Schummers2007} \\
\hline
\multicolumn{4}{l}{{\em Intrinsic (voltage gated) currents}}\\
$g_{Na}$ & $17.9$~$\mu$S & Sodium current, peak conductance & \cite{Marino2005} \\
$E_{Na}$ & $50$~mV & Sodium current, reversal potential & \cite{Marino2005} \\
$l_{Na}$ & 3 & Sodium current, no.\ of activation sites & \cite{Marino2005} \\
$k_{Na}$ & 1 & Sodium current, no.\ of inactivation sites & \cite{Marino2005} \\
$g_{Kd}$ & $3.46$~$\mu$S & Potassium current, peak conductance & \cite{Marino2005} \\
$E_{Kd}$ & $-90$~mV & Potassium current, reversal potential & \cite{Marino2005} \\
$l_{Kd}$ & 4 & Potassium current, no.\ of activation sites & \cite{Marino2005} \\
$k_{Kd}$ & 0 & Potassium current, no.\ of inactivation sites & \cite{Marino2005} \\
$g_{M,E}$ & $279$~nS & M-channel, peak conductance, excit.\ neurons& \cite{Marino2005} \\
$g_{M,I}$ & $27.9$~nS & M-channel, peak conductance, inhib.\ neurons& \cite{Schummers2007} \\
$E_{M}$ & $-85$~mV & M-channel, reversal potential & \cite{Marino2005} \\
$l_{M}$ & 1 & M-channel, no.\ of activation sites & \cite{Marino2005} \\
$k_{M}$ & 0 & M-channel, no.\ of inactivation sites & \cite{Marino2005}\\
\hline
\multicolumn{4}{l}{{\em Background currents}}\\
$E_{bgE}$ & $-5$~mS & Excitatory current, reversal potential & \cite{Schummers2007} \\
$E_{bgI}$ & $-70$~mV & Inhibitory current, reversal potential & \cite{Schummers2007} \\
$\tau_{bgE}$ & $2.7$~ms & Excitatory current, time constant & \cite{Schummers2007} \\
$\tau_{bgI}$ & $10.7$~ms & Inhibitory current, time constant & \cite{Schummers2007} \\
$\bar{g}_{bgEE}$ & $8.79$~nS & average excit.\ to excit.\ conductance & \cite{Schummers2007} \\
$\bar{g}_{bgEI}$ & $28.8$~nS & average inhib.\ to excit.\ conductance & \cite{Schummers2007} \\
$\bar{g}_{bgIE}$ & $17.5$~nS & average excit.\ to inhib.\ conductance & \cite{Schummers2007} \\
$\bar{g}_{bgII}$ & $57.6$~nS & average inhib.\ to inhib.\ conductance & \cite{Schummers2007} \\
$\sigma_{bgE}$ & $0.157$~nS & Noise strength, excit.\ conductance & \cite{Schummers2007} \\
$\sigma_{bgI}$ & $0.313$~nS & Noise strength, inhib.\ conductance & \cite{Schummers2007} \\
\end{tabular}
\begin{flushleft}
\end{flushleft}
\label{tab:neuron}
\end{table}

\paragraph*{Intrinsic currents}
The Hodgkin-Huxley-type neuron model implements three intrinsic voltage-gated currents $I_{int,X} =
I_{Na} + I_{Kd} + I_{M,X}$, a fast sodium
current $I_{Na}$, a delayed-rectified potassium current $I_{Kd}$, and a slow
non-inactivating population specific potassium current $I_{M,X}$. The
intrinsic currents of each neuron follow:
\begin{equation}
I_Z = \bar{g}_Z m_{Zact}^{l_Z}(V) m_{Zinac}^{k_Z}(V)(V-E_Z), \qquad Z\in \{Na,Kd,M\},
\end{equation}
with $\bar{g}_Z$ the peak conductance, $m_{Zact}$ and $m_{Zinact}$ the activation and inactivation variables, $l_Z$ and $k_Z$ the number of activation and inactivation sites, and $E_Z$ the reversal potential of the
channel. The
peak conductance for the $M$-current is population selective $\bar{g}_{M,X}$, to account for weaker adaptation in inhibitory neurons. The
dynamics of activation ($D=act$) and inactivation ($D=inac$) are given by:
\begin{eqnarray*}
\frac{d m_{ZD}}{dt} &=& \alpha_{ZDo} (1 - m_{ZD}) - \alpha_{ZDc} m_{ZD}\\
\alpha_{ZDd} &=& \frac{v_1 v_2 (V)}{\exp(v_3(V)) + v_4}\\
\end{eqnarray*}
with $\alpha_{ZDd}$ opening ($d=o$) and closing ($d=c$) transition rates. The
kinetics follow \cite{Destexhe1999} and are summarized in
Table~\ref{tab:vg_kinetics}
\begin{table}[ht!]
\caption{
\bf{Channel dynamics}}
\begin{tabular}{lcccc}
Gating var. & $v_1$ [mV$^{-1}$] & $v_2(V)$ [mV] & $v_3(V)$ [mV] & $v_4$\\ 
\hline
$\alpha_{Na\:act\:o}$ & $0.32$ & $-(V + 45)$ & $-(V + 45)/4$ & $-1$ \\
$\alpha_{Na\:inac\:o}$ & $0.128$ & $1$ & $(V + 51)/18$ &  $0$ \\
$\alpha_{Kd\:act\:o}$ & $0.032$ & $- (V + 40)$ & $- (V +40)/5$ &  $-1$ \\
$\alpha_{M\:act\:o}$ & $2.9529 \times 10^{-4}$ & $- (V + 30)$ & $ - (V + 30)/9$ & $-1$ \\
$\alpha_{Na\:act\:c}$ & $0.28$ & $V + 18$ & $(V + 18)/5$ & $-1$ \\
$\alpha_{Na\:inact\:c}$ & $4.$ & $1$ & $- (V+28)/5$ & $1$ \\
$\alpha_{Kd\:act\:c}$ & $0.5$ & $1$ & $(V+45)/40 $ & $0$ \\
$\alpha_{M\:act\:c}$ & $2.9529 \times 10^{-4}$ & $V+30$ & $(V+30)/9$ & $-1$ \\
\hline
\end{tabular}
\begin{flushleft}Expressions for channel dynamics as in \cite{Destexhe1999}
\end{flushleft}
\label{tab:vg_kinetics}
\end{table}

\paragraph*{Synaptic currents}  Each neuron receives a set of (lateral \&
afferent) glutamatergic and lateral GABA-ergic synaptic
currents.
\[
 I_{syn,X}  = \frac{1}{N_{Aff}} \sum_{j} I_{AMPAaff,X}^j + \frac{1}{N_{Xe}} \sum_k \left( I_{AMPA,X}^k + I_{NMDA,X}^k \right) + \frac{1}{N_{ci}} \sum_m I_{GABA_A}^m 
\]
 with $N_{Aff},N_{Xe},N_{ci}$ the number of received connections, and $j,k,m$
 the indices of projecting afferent, excitatory, and inhibitory neurons with
 $X \in \{E,I\}$ the target population. The current through the specific
 receptor type at every synapse is governed by the introduced receptor dynamics. The post-synaptic current $I_{R}$ with $R \in \{AMPA,NMDA,GABA_A\}$ is given by:
\[ 
I_{R} = \underbrace{\bar{g}_{R} B^{l_R} \left(\Sigma O_{R}\right)}_{g_{R}} (V - E_R ), \qquad l_R = \left\{ \begin{array}{ll}
 1 & \mbox{for R  = NMDA} \\
 0 & \mbox{otherwise} \end{array} \right. ,
\]
with $\bar{g}_{R}$ the receptor specific
peak conductance, $\Sigma O_R$ the sum of open states, $V$ the membrane potential of the post-synaptic neuron, and $E_R$ the reversal potential. NMDA-receptors also express a voltage and magnesium dependent block $B =
\left( 1+ \exp \left( -0.062 V + 1.2726   \right) M\!g \right)^{-1} $ and
$M\!g$ the extracellular magnesium concentration \cite{Jahr1990}. For the considerations of tuning in the conductances the excitatory synaptic conductances $g_{NMDA}$ and $g_{AMPA}$ are aggregated ($g_e = g_{NMDA} + g_{AMPA}$). For the inhibitory conductances the experimental limitations to distinguish between synaptic inhibitory conductances and adapting intrinsic conductances we combine the GABA$_A$ and the slow non-inactivating potassium current. ($g_i = g_{GABA_A} + g_{M,E}$) to provide compatible values \cite{Schummers2007}.

\paragraph*{Synaptic peak conductances} A major difference to earlier models
of V1 (cf. \cite{Stimberg2009} and \cite{Roy2013arxiv}) are the detailed
synaptic kinetics. Therefore, afferent and lateral peak conductances had to be
re-adjusted. To determine the peak-conductances for afferent ($\bar{g}_{AMPAaff,E}$ and
$\bar{g}_{AMPAaff,I}$) and inhibitory synapses ($\bar{g}_{GABA_A}$) we
stimulated simple exponential AMPA and GABA$_A$ synapses parametrized as in \cite{Stimberg2009} and
the introduced detailed ones with the same $40$~Hz Poisson spike-trains for 2~s. Then we
determined the peak conductance for which the average conductances were equal.

For the excitatory lateral peak conductances we assumed a 4:1 ratio for AMPA to
NMDA receptors and we followed the paths described in
\cite{Stimberg2009} and in \cite{Roy2013arxiv} to determine peak
conductance values for connections to excitatory and inhibitory neurons. In
detail we derived the peak conductances ($\bar{g}_{AMPA,E}$ and
$\bar{g}_{NMDA,E}$, and $\bar{g}_{AMPA,I}$, $\bar{g}_{NMDA,I}$) for the
pinwheel-domain model by exactly following the procedure described in \cite{Stimberg2009} of
matching orientation selectivity indices (OSI) for different mapOSIs to the
data by \cite{Marino2005} (for definition of OSI and mapOSI see below). For the salt-and-pepper network we matched the peak
conductances by using Kolmogorov-Smirnov-tests to compare OSI-distributions
with the data by \cite{Runyan2013}, as described in \cite{Roy2013arxiv}. In
both cases we used a grid search in the space spanned by peak conductances to excitatory and
inhibitory neurons and selected the best matching point. Parameter used for
peak-conductances are comprised in Tab.~\ref{tab:currents}.

\begin{table}[!ht]
\caption{
\bf{Ligand gated receptors and currents}}
\begin{tabular}{llll}
Parameter & Value & Description & Source \\ 
\hline
\multicolumn{4}{l}{{\em Synaptic -- Currents}}\\
$\bar{g}_{AMPAaff,E}$ & $549.51$~nS  & Peak aff. AMPAR cond. to
exc. neurons & * \\
$\bar{g}_{AMPAaff,I}$ & $0.73\bar{g}_{AMPAaff,E}$  & Peak aff. AMPAR cond. to
inh. neurons & \cite{Stimberg2009} \\
$\bar{g}_{GABA_A}$ & $281.8$~nS & Peak rec. GABA$_A$R conductance & * \\
$E_{AMPA}$ & $0$~mV & AMPAR reversal potential & \cite{Destexhe1998} \\
$E_{NMDA}$ & $0$~mV & NMDAR reversal potential & \cite{Destexhe1998} \\
$E_{GABA_A}$ & $-70$~mV & GABA$_A$ reversal potential & \cite{Destexhe1998} \\
$Mg$ & $1$~mM/M & Unit-free magnesium concentration & \cite{Jahr1990} \\
\multicolumn{4}{l}{{\em pinwheel-domain specific}}\\
$\bar{g}_{AMPA,E}$ & $879.40$~nS & Peak rec. AMPAR conductance  to
exc. & * \\
$\bar{g}_{AMPA,I}$ & $1538.61$~nS & Peak rec. AMPAR conductance  to
inh. & * \\
$\bar{g}_{NMDA,E}$ & $219.80$~nS & Peak rec. NMDAR conductance  to
exc. & * \\
$\bar{g}_{NMDA,I}$ & $384.65$~nS & Peak rec. NMDAR conductance  to
inh. & * \\
\multicolumn{4}{l}{{\em salt-and-pepper specific}}\\
$\bar{g}_{AMPA,E}$ & $659.40$~nS & Peak rec. AMPAR conductance  to
exc. & * \\
$\bar{g}_{AMPA,I}$ & $879.20$~nS & Peak rec. AMPAR conductance  to
inh. & * \\
$\bar{g}_{NMDA,E}$ & $164.84$~nS & Peak rec. NMDAR conductance  to
exc. & * \\
$\bar{g}_{NMDA,I}$ & $219.80$~nS & Peak rec. NMDAR conductance  to
inh. & * \\

\hline
\multicolumn{4}{l}{{\em Extra-synaptic}}\\
$\bar{g}_{amb}$ & $2.6$~nS & Peak extra-synaptic NMDAR conductance &  \cite{Bentzen2009}\\
$\alpha$ & $0.54$~M$^{-H}$ & Factor based on transition rates & \cite{Bentzen2009}\\
$H$ & $1.5$ & Hill-coefficient of extra-synaptic NMDAR & \cite{Bentzen2009} \\
$E_{som}$ & $55$~mV & Extra-synaptic NMDAR reversal potential & \cite{Bentzen2009} \\
$G_{amb,E}$ & $0$--$2.5$~$\mu$M & Amb. Glut. concentr. to exc. neurons & \cite{Herman2007}$^\circ$ \\
$G_{amb,E}$ & $0$--$2.5$~$\mu$M & Amb. Glut. concentr. to inh. neurons & \cite{Herman2007}$^\circ$ \\
\hline
\end{tabular}
\begin{flushleft} * Peak synaptic conductances were determined as described in
  synaptic peak conductance paragraph. The ambient glutamate concentration are
  varied in a biological plausible range (cf. \cite{Herman2007})
\end{flushleft}
\label{tab:currents}
\end{table}

\paragraph*{Extrasynaptic ligand-gated currents} Extrasynaptic NMDA-receptors
are activated by ambient glutamate $G_{amb}$. Different densities of
NMDA-receptors on excitatory and inhibitory neurons provide population
specific currents $I_{amb,X}$.  The currents follow the steady
state descriptions of extrasynaptic NMDA-receptors (eNMDAR) in \cite{Bentzen2009}:
\begin{eqnarray}
 I_{amb,X} &=& \bar{g}_{amb} B [eNMDAo] \left(V - E_{som} \right)
 \nonumber 
\end{eqnarray}
with $\bar{g}_{amb}$ the peak conductance of eNMDARs, $B$
the dynamics of its magnesium block , $[eNMDAo]= \alpha
G_{amb}^H$ the fraction of open eNMDARs following a power law-dependence on
ambient glutamate, and $E_{som}$ the Nernst-potential of the
eNMDARs (Parameters in Tab.~\ref{tab:currents}). 

\paragraph*{Background currents}
We consider the network to be embedded in surrounding neuronal
activity. Therefore, each neuron receives synaptic background
activity: 
\[
I_{bg,X} = g_{bgXE}(t)(V - E_{bgE}) - g_{bgXI} (V - E_{bgI})
\]
with specific reversal potentials ($E_{bgE}$,$E_{bgI}$) and
fluctuating conductances $g_{bgXY}, \;\; Y \in \{E,I\}$ the source population, following an
  Ornstein-Uhlenbeck process:
\[
 d g_{bgXY} = \tau_{bgY} ( \bar{g}_{bgXY} - g_{bgXY}) + \sigma_{bgY} d W \, ,
\]
with $\tau_{bgY}$ the mean reversion speed, $\bar{g}_{bgXY}$ the average
background conductance and $\sigma_{bgY}$ the noise strength.

\begin{figure}[!ht]
\begin{center}
\includegraphics[width=5in]{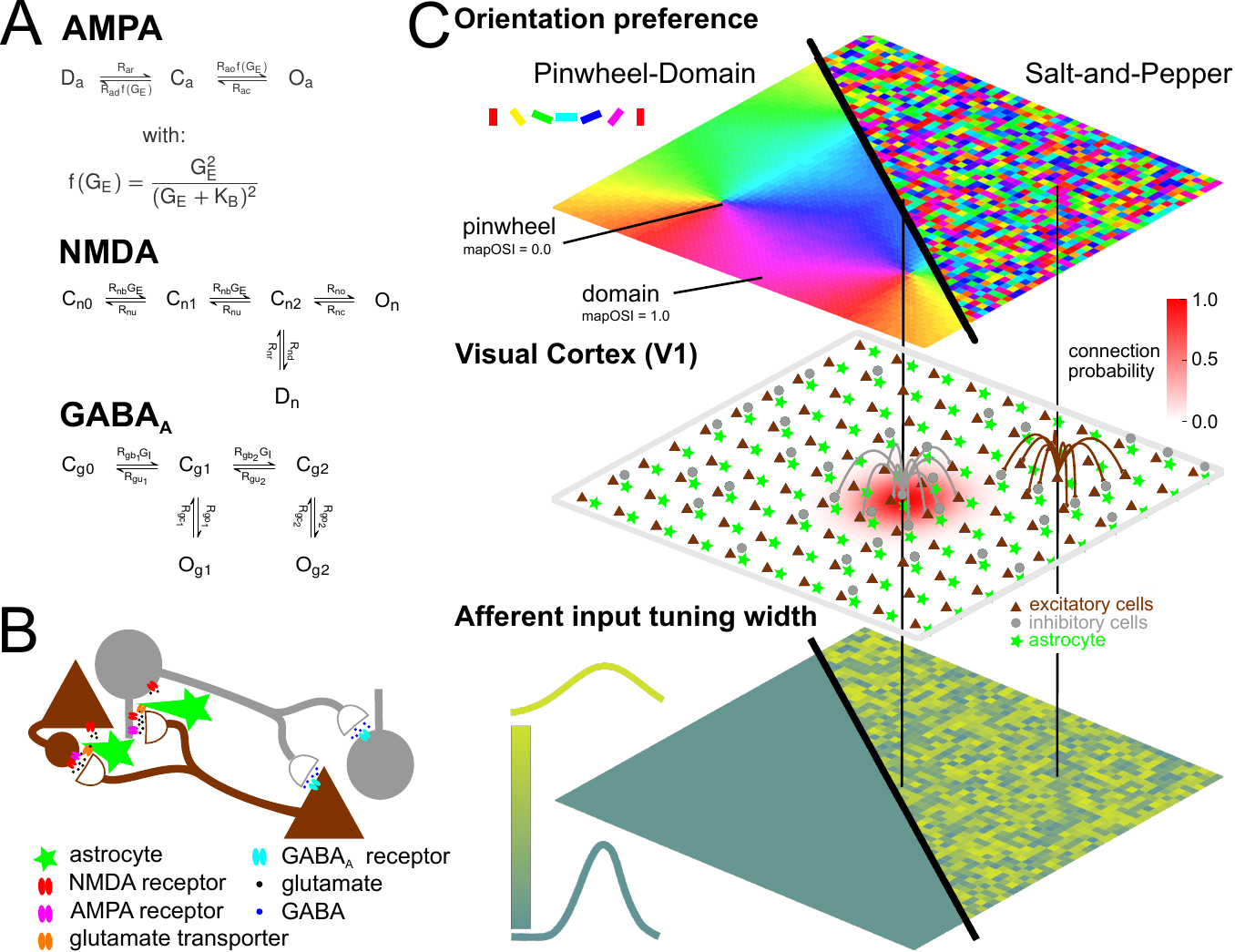}
\end{center}
\caption{
{\bf V1 network model.  A} Synapses use explicit neurotransmitter descriptions $G_E$
and $G_I$, which activate receptors described by extended kinetic schemes,
with several closed $C$, desensitized $D$ and open $O$ stages and constant $R$ and
transmitter dependent $Rf(G)$ transition rates. AMPA follows the description in
\cite{Saftenku2005}, NMDA based on \cite{Lester1992}, and GABA$_A$ on
\cite{Destexhe1998}. {\bf B} The model contains glutamatergic connections to inhibitory
and to excitatory neurons, with NMDA \& AMPA receptors. Effect of glutamate
transporters on the glutamate time course in the two types of connection is
separately varied due to the difference in synapse geometry. Extrasynaptic
NMDA-receptors are activated by ambient glutamate. {\bf C} The two one-layered V1 network models are composed of excitatory (brown)
and inhibitory (gray) neurons which receive tuned excitatory affentent and
lateral inhibitory and excitatory inputs. Neurons a the same location in the
network share the preferred orientation, which is either organized in a
pinwheel-domain (left) or salt-and-pepper map (right). Lateral connections are drawn from a 2d-Gaussian
independent of preferred orientation. Afferent input already carries some
tuning. For the pinwheel-domain model every neuron receives input from equally
tuned neurons as in \cite{Stimberg2009}. For the salt-and-pepper map afferent
tuning width is sampled from independent distributions for exc. and inhibitory
neurons. }
\label{fig:model}
\end{figure}

\paragraph*{V1 network} The network layout is similar to \cite{Stimberg2009} and to \cite{Roy2013arxiv}. Two populations, an excitatory (size: $N_E$) and an inhibitory (size:
$N_I$) population of neurons represent layer 2-3 of the primary visual cortex, on
the one hand for species with a pinwheel-domain organization, e.g., ferrets
(Fig.~\ref{fig:model}C left part), on the other hand for species with
a salt-and-pepper organization, e.g., mice (Fig.~\ref{fig:model}C right part).
For both models the excitatory neurons are regularly placed on a 2d-grid of size $\sqrt{N_E} \times
\sqrt{N_E}$. Inhibitory neurons are randomly placed on a third of all grid points. Each neuron receives a number ($N_{aff}$) of afferent Poisson inputs with stimulus
specific rates. Additionally, excitatory and inhibitory neurons receive fixed
specific numbers of recurrent excitatory ($N_{ee}$, $N_{ie}$) inputs and a number of recurrent inhibitory ($N_{ci}$) inputs. The model for mouse
features lower numbers of excitatory connections than the one for
ferret (cf. Table~\ref{tab:geometry}). Independent of species all recurrent connections are randomly drawn,
from the same radial symmetric 2d-Gaussian distance distribution, using the algorithm proposed in
\cite{Efraimidis2008},
\[
P(r)=\left\{ \begin{array}{ll}
 0 & \mbox{for $r=0$ (no self-connections)};\\
  1/\sqrt{2\pi\sigma}\exp(-r^2/{2\sigma_C^2}) & \mbox{otherwise,} \end{array} \right.
\]   
with $r$ the distance (in gridpoints) to the presynaptic neuron, and $\sigma_C$ the width of the
Gaussian. Thereby connections between neighboring neurons are more likely
(see Figure~\ref{fig:model}A). Each individual connection gets a transmission delay, which comprise
synaptic and conduction delays, and is drawn from a
gamma distribution $\Gamma (k_Y,\theta_Y)$ with shape $k_Y$ and scale
$\theta_Y$ parameters specific for the source population $Y \in
\{E,I\}$. Connections at the boundaries are generated using periodic boundary
conditions. Parameters can be found in Table~\ref{tab:geometry}.


\begin{table}[!ht]
\caption{
\bf{Geometry and stimulation parameters}}
\begin{tabular}{llll}
Parameter & Value & Description & Source \\ 
\hline
\multicolumn{4}{l}{{\em Geometry}}\\
$N_E$ & 2500 & Number of excit. neurons & \cite{Stimberg2009} \\
$N_I$ & 833 & Number of inhib. neurons & \cite{Stimberg2009} \\
$N_{Aff}$ & 20 & Number of afferent inputs & \cite{Stimberg2009} \\
$\sigma_C$ & 4 & Spread of recurrent conn. (std. dev.) & \cite{Stimberg2009}\\
$k_E$ & $7$ & Shape of Gamma distribution exc. conn. & \cite{Stimberg2009}*\\
$\theta_E$ & $0.6$ & Scale of Gamma distribution exc. conn. & \cite{Stimberg2009}*\\
$k_I$ & $2.5$ & Shape of Gamma distribution inh. conn. & \cite{Stimberg2009}*\\
$\theta_I$ & $0.6$ & Scale of Gamma distribution inh. conn. & \cite{Stimberg2009}*\\
\multicolumn{4}{l}{{\em ferret specific}}\\
$N_{ee}$ = $N_{ie}$ & 100 & Number of excit. recurrent inputs& \cite{Stimberg2009}\\
$N_{ci}$ & 50 & Number of inhib. recurrent inputs & \cite{Stimberg2009}\\
\multicolumn{4}{l}{{\em mouse specific}}\\
$N_{ee}$  & 25 & Number of excit. to excit. inputs&
\cite{Roy2013arxiv}\\
$N_{ie}$ & 50 & Number of excit. to inhib. inputs& \cite{Roy2013arxiv}\\
$N_{ci}$ & 50 & Number of inhib. recurrent inputs & \cite{Roy2013arxiv}\\
\hline
\multicolumn{4}{l}{{\em Stimulation}}\\
$f_{A,max}$ & $30$~Hz & max afferent firing rate & \cite{Stimberg2009} \\
$r_{base}$ & $0.1$ & fraction of stimulus indep. rate & \cite{Stimberg2009}
\\
\multicolumn{4}{l}{{\em ferret specific}}\\
$w_A$ & $27.5$~deg. & input tuning width & \cite{Stimberg2009} \\
\multicolumn{4}{l}{{\em mouse specific}}\\
$w_{EA}$ & $17.5$~deg. & input tuning width & \cite{Roy2013arxiv} \\
$w_{IA}$ & $57.5$~deg. & input tuning width & \cite{Roy2013arxiv} \\
$\sigma_{wEA}$ & $16$~deg. & input tuning width & \cite{Roy2013arxiv} \\
$\sigma_{wIA}$ & $48$~deg. & input tuning width & \cite{Roy2013arxiv} \\
\end{tabular}
\begin{flushleft} *~Mean and standard deviation of the gamma distribution are
  matched to the values in \cite{Stimberg2009}
\end{flushleft}
\label{tab:geometry}
\end{table}

\paragraph*{Organization of preferred orientations}
For species which express a pinwheel-domain organization we generate the
preferred orientation $\theta(x,y)$ for each neuron based on its location $(x,y)$
within a pinwheel-domain map representing 4 pinwheels (see
Figure~\ref{fig:model}C left part). The map is constructed by, first producing a single pinwheel in
the first quadrant $1q$ using equally spaced coordinates $-1\leq x<1$ and $-1\leq
y<1$ and deriving every neurons preferred angle by:
\[
\theta_{1q}(x,y) =  \frac{90}{\pi} \textnormal{atan2}(x,y),
\]
second the full map is generated by mirroring the first into the other three
quadrants. 
For species without a distinct organization we generate a {\em
  salt-and-pepper}-map by uniformly distributing preferred orientations
randomly (see Fig.~\ref{fig:model}C right part).

\paragraph*{Stimulation} 
All neurons receive a number $N_{Aff}$ of individual afferent
Poisson-inputs generated by a full-field stimulation with a static
fixed orientation $\theta_{stim}$. The neuron specific rate
\[
f_A \left( \theta_{stim}, \theta(x,y), w_{XA}(x,y) \right) = f_{A,max} \left( r_{base} +
\left(1-r_{base}\right) \exp \left(-\frac{\left(\theta_{stim} - \theta(x,y)
  \right)^2}{4\sigma_{XA}^2(x,y)}\right) \right),
\]
depends on: selected stimulus orientation $\theta_{stim}$, preferred
orientation of the neuron $\theta(x,y)$, a base-line firing rate $r_{base}$, the
maximal firing rate for optimal stimulation $f_{A,max}$, and the population
($X \in \lbrace E,I \rbrace $)
and neuron specific input tuning width $w_{XA}(x,y)$. For ferret all neurons
independent of population receive input with identical tuning width
$w_{XA}(x,y) = w_A$ (Fig.~\ref{fig:model}C left part). For mouse individual neuron receive specific
afferent inputs drawn from two truncated (0-90) Gaussian-distributions
differently parametrized for excitatory (mean: $w_{EA}$, standard deviation
$\sigma_{wEA}$) and inhibitory (mean: $w_{IA}$, standard deviation
$\sigma_{wIA}$) neurons, cf. Fig.~\ref{fig:model}C right part,
\cite{Roy2013arxiv}, and Table~\ref{tab:geometry}

\paragraph*{Blocking glutamate transport} For the network we explore two ways
how blocking glutamate transport affects tuning. First, at the synapses
different decay times for $\tau_{fEE}$ and $\tau_{fIE}$ are considered. Second
different NMDA-receptor current strengths to excitatory and inhibitory neurons
are considered. We allow different $G_{amb,E}$ and $G_{amb,I}$, as a proxy for
different densities of NMDA-receptors. 

\paragraph*{Numerical Simulations}
The synapses and the network model are implemented in Python 2.7 using Brian2 to generate C++
code. We used the Euler-integration scheme provided by the toolbox with an
integration step of $0.01$~ms. Every simulation is run first for $400$~ms as
initialization phase without recording data, then for $1600$~ms data is recorded.

\paragraph*{Analyses}
Two different measures were used to analyze orientation tuning. First the
orientation selectivity index (OSI; \cite{Swindale1998}), given by

\[
\textnormal{OSI} =  \left. \sqrt{\left( \sum_i R(\theta_i) \cos(2\theta_i) \right)^2
  + \left( \sum_i R(\theta_i) \sin(2\theta_i) \right)^2 }
  \right/ \sum_i R(\theta_i),
\]  
$R(\theta_i)$ is the investigated quantity (e.g. firing rate) observed as
response to a stimulation with orientation $\theta_i$. Stimulation
orientations $\theta_i$ have to span the entire range of possible orientations
and have to be equally spaced. Values for OSI range from 0 (unselective) to
1 (highly selective). When using the number of sites with a specific orientation preference of neighboring
pixels in a radius of 8 pixels instead of $R(\theta_i)$ the OSI-measure can be
used to derive the {\em mapOSI}, which quantifies the homogeneity of lateral inputs
(cf. Fig~\ref{fig:model}C). As second measure we used the half-width at half
max (HWHM) of the response tuning curves. 

Instead of deriving tuning curves for each individual neuron by stimulating
with different orientations we generate pseudo-neurons from a single
simulation with one fixed stimulus orientation $\theta
=43.8$~deg. Pseudo-neurons are generated by splitting all excitatory neurons
into batches of 50 neurons based on their mapOSI (in
pinwheel-domain case) or afferent input tuning width (in salt-and-pepper
case). E.g., the 50 neurons with the smallest mapOSI constitute a pseudo
neuron. As the 50 neurons have different preferred orientations we can
consider these as stimulations with different offsets to the preferred
orientation of a pseudo-neuron. Therefore,
neurons (of a range of mapOSIs or afferent tuning width) with a preferred
orientation close to $\theta(x,y) = 43.8$ will give the
response of the pseudo-neuron (with the mapOSI or afferent tuning width)
stimulated close to its preferred orientation. Equally spaced
stimulations of the pseudo-neuron are obtained by, first fitting a flat-topped von-Mises distributions
\cite{Swindale1998} to the pseudo-neuron data, and second selecting points
with a $10$~deg. difference from the obtained distributions. For the
separation in pseudo-neurons close to pinwheel and within domains we used
mapOSI $\leq 0.4$ and $0.6 < $ mapOSI $\leq 0.9$, respectively.

\section*{Acknowledgments}
The authors like to acknowledge the support team of Brian 2, which rapidly
developed requested features which made it possible to run the simulations in
standalone c++ code. 

\bibliography{library}

\end{document}